\documentclass[aps,twocolumn,showpacs,floatfix]{revtex4}
\usepackage{amsmath}
\usepackage{amsfonts}
\usepackage{amssymb}
\usepackage{graphicx}
\begin{document}
\title{Electromagnetically Induced Transparency from Two Phonon Processes in Quadratically Coupled Membranes}
\author{Sumei Huang and G. S. Agarwal}
\affiliation{Department of Physics, Oklahoma State University,
Stillwater, Oklahoma 74078, USA}
\date{\today}
\begin{abstract}
We describe how electromagnetically induced transparency can arise in quadratically coupled optomechanical systems. Due to quadratic coupling the underlying optical process involves a two phonon process in optomechanical system and this two phonon process makes the mean amplitude, which plays the role of atomic coherence in traditional EIT, zero. We show how the fluctuation in displacement can play a role similar to atomic coherence and can lead to EIT-like effects in quadratically coupled optomechanical systems. We show how such effects can be studied using the existing optomechanical systems.
\end{abstract}
\maketitle
\section{Introduction}

 The radiation pressure coupling between the nano mirror and the radiation field is known to depend on the displacement of the mirror via the cavity frequency \cite{Meystre}. This coupling can depend linearly or quadratically on the displacement depending on the location of the mirror with respect to nodes and antinodes of the cavity modes. The case most extensively discussed in the literature corresponds to placing the mirror at a node so that the coupling is linear in displacement \cite{Anetsberger,Aspelmeyer,Dobrindt,Marquardt,Hartmann,Hammerer,Bhattacharya,Paternostro}. Nanomechanical systems with linear reactive coupling have also been studied \cite{Tang,Clerk,Sumei}. The case of quadratic coupling has not been studied that extensively as the coupling is generally small. However recent works \cite{Sankey,Thompson,Jayich} have shown a way to get much larger quadratic couplings and therefore one should study the novel consequences of quadratic coupling in detail. The quadratic coupling in phonon picture implies two phonon processes and such couplings in analogy to well known quantum optical Hamiltonians \cite{Walls} naturally lead to the possibility of squeezing the mechanical oscillator \cite{Rai,Nunnenkamp,Rae}. The question that we examine in this paper is how to probe the effects of such two phonon processes by using pump and probe fields of respective frequencies $\omega_{c}$ and $\omega_{p}$. We expect that the two phonon processes should show up when the frequency difference $\omega_{p}-\omega_{c}$ is about 2$\omega_{m}$ where $\omega_{m}$ is the frequency of the mechanical oscillator. At the outset we want to mention that in case of single phonon processes (linear coupling); the mean amplitude of the oscillator is nonzero and it leads to the modulation of the output fields whereas for two phonon processes the mean response of the oscillator is zero \cite{Meystre} and thus any modulation of the output fields has to come from mean values of $x^2$ which is a temperature dependent quantity. We further reveal the possibility of an analog of electromagnetically induced transparency arising from temperature dependent oscillators mean potential energy. This is different from the linear coupling case where the mean amplitude of the oscillator determines the EIT behavior \cite{Agarwal,Kippenberg,Vahala}. For our case of two phonon processes the role of atomic coherence in traditional EIT is played by the mean value of $x^2$ which in addition to temperature also depends on the strength of the coupling field.

The paper is organized as follows. In Sec. II, we describe the model under study, give the equation of motion for the system operators, obtain the output field at the probe frequency. In Sec. III, we discuss the effect of the quadratic optomechanical coupling on the output field at the probe frequency. We find that the EIT-like dip appears in the output field at the probe frequency.
\section{Model}
\begin{figure}[!h]
\begin{center}
\scalebox{0.8}{\includegraphics{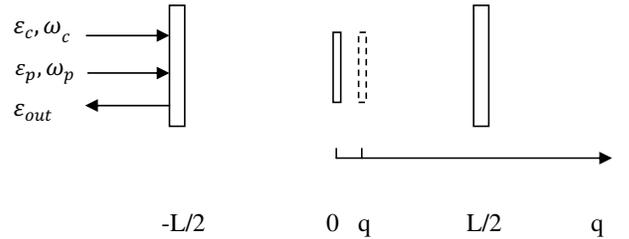}} \caption{\label{Fig1} Sketch of the studied system. A strong coupling field at frequency $\omega_{c}$ and a weak probe field at frequency $\omega_{p}$ are injected into the cavity through the left mirror. A membrane with finite reflectivity is located at the middle position of the cavity. After the interaction between the cavity field and the membrane, the output field will contain three frequencies ($\omega_{c}$, $\omega_{p}$, $2\omega_{c}-\omega_{p}$).}
\end{center}
\end{figure}
Let us start with a sketch of the system as shown in Fig. 1 \cite{Thompson,Jayich}. A membrane with finite reflectivity $r_{c}$ is placed inside the cavity formed by two fixed mirrors separated from each other by a distance $L$. A strong coupling field of amplitude $\varepsilon_{c}$ and a weak probe field of amplitude $\varepsilon_{p}$ are sent into the cavity through the partially transmitting left mirror, the right mirror is perfectly reflecting. The membrane with mass $m$ is assumed to be located at an extremum of the frequency $\omega_{0}$ of the cavity field, thus the cavity is quadratically coupled to the displacement of the membrane, and we denote the quadratic coupling constant by $g$. Moreover, the membrane contacts with the environment in thermal equilibrium at temperature $T$. Hence the system's Hamiltonian takes the form
\begin{eqnarray}\label{1}
H&=&\hbar\omega_{0}c^{\dag}c+\hbar gc^{\dag}cq^{2}+\frac{p^{2}}{2m}+\frac{1}{2}m\omega_{m}^{2}q^2\nonumber\\& &+i\hbar \varepsilon_{c}(c^{\dag}e^{-i\omega_{c}t}-ce^{i\omega_{c}t})+i\hbar (\varepsilon_{p}c^{\dag}e^{-i\omega_{p}t}-\varepsilon_{p}^{*}ce^{i\omega_{p}t}),\nonumber\\
\end{eqnarray}
in which $c$ and $c^{\dag}$ denote the annihilation and creation operators of the cavity, while $q$ and $p$ are the position and momentum operators of the membrane. $\varepsilon_{c}$ and $\varepsilon_{p}$ are defined by $\varepsilon_{c}=\sqrt{2\kappa \wp_{c}/(\hbar\omega_{c})}$ and
$\varepsilon_{p}=\sqrt{2\kappa \wp_{p}/(\hbar\omega_{p})}$, where
$\wp_{c}$ is the power of the coupling field, $\wp_{p}$ is the power of the probe field, and $\kappa$ is the cavity decay rate.
In the rotating frame at the frequency $\omega_{c}$ of the coupling field, $c(t)=\tilde{c}(t)e^{-i\omega_{c}t}$, using the Heisenberg equation of motion and adding the corresponding noise and damping terms, we can obtain the equation of motion for the mirror and the cavity variables.
\begin{eqnarray}\label{2}
\frac{dq}{dt}&=&\frac{p}{m},\nonumber\\
\frac{dp}{dt}&=&-m\omega_{m}^{2}q-2\hbar g\tilde{c}^{\dag}\tilde{c}q-\gamma_{m}p+\xi,\nonumber\\
\frac{d\tilde{c}}{dt}&=&-[\kappa+i(\omega_{0}-\omega_{c}+gq^{2})]\tilde{c}+\varepsilon_{c}+\varepsilon_{p}e^{-i(\omega_{p}-\omega_{c})t}\nonumber\\
& &+\sqrt{2\kappa}\tilde{c}_{in},\nonumber\\
\frac{d\tilde{c}^{\dag}}{dt}&=&-[\kappa-i(\omega_{0}-\omega_{c}+gq^{2})]\tilde{c}^{\dag}+\varepsilon_{c}+\varepsilon_{p}^{*}e^{i(\omega_{p}-\omega_{c})t}
\nonumber\\& &+\sqrt{2\kappa}\tilde{c}_{in}^{\dag},\nonumber\\
\end{eqnarray}
in which $\gamma_{m}$ is the damping rate of the membrane, $\xi$ is the Langevin force from the environment with zero mean value, and $\tilde{c}_{in}$ is the input vacuum noise with zero mean value. From Eq. (2), we can obtain the expectation values of the system operators at the steady state. These are
\begin{eqnarray}\label{3}
q_{0}=0, p_{0}=0, c_{0}=\frac{\varepsilon_{c}}{\kappa+i(\omega_{0}-\omega_{c})},
\end{eqnarray}
where from now on we drop the tilde from $\tilde{c}_{0}$. It is seen that at the steady state, the membrane's displacement is zero, and the amplitude $c_{0}$ of the cavity field is unrelated to the position of the membrane so that the output field is not modified by the mean amplitude of the membrane, which is different from that in the linear coupling case. We note that in the case of linear coupling of the membrane the mean value of the displacement plays the same role as atomic coherence in case of EIT with atomic vapors. Now such a coherence term is zero and hence a key element for the occurrence of EIT for quadratically coupled membrane is zero. We propose here a way out of this difficulty. Since mean value of $q$ is zero however its variance which is proportional to potential energy of the membrane is expected to be nonzero. Thus in our proposal for EIT with quadratically coupled optomechanical systems the quantity $\langle q^{2}\rangle$ will be central. This peculiarity is related to the fact that the underlying physical process is a two phonon process. Thus in the following we turn to calculate the evolutions of the expectation values of $q^{2}$, $p^{2}$, and $qp+pq$, which can be obtained with the help of Eq. (2) and the factorization assumption $\langle a b c\rangle=\langle a \rangle \langle b\rangle \langle c\rangle$. Using the same method, we also can obtain the evolutions of the expectation values of $c$ and $c^{\dag}$. Hence the complete set of underlying equations for our system would be

\begin{eqnarray}\label{4}
\frac{d}{dt}\langle c\rangle&=&-[\kappa+i(\omega_{0}-\omega_{c}+g\langle q^{2}\rangle)]\langle c\rangle+\varepsilon_{c}\nonumber\\& &+\varepsilon_{p}e^{-i(\omega_{p}-\omega_{c})t},\nonumber\\
\frac{d}{dt}\langle c^{\dag}\rangle&=&-[\kappa-i(\omega_{0}-\omega_{c}+g\langle q^{2}\rangle)]\langle c^{\dag}\rangle+\varepsilon_{c}\nonumber\\& &+\varepsilon_{p}^{*}e^{i(\omega_{p}-\omega_{c})t},\nonumber\\
\frac{d}{dt}\langle q^2\rangle&=&\frac{1}{m}\langle pq+qp\rangle,\nonumber\\
\frac{d}{dt}\langle p^2\rangle&=&[-m\omega_{m}^2-2\hbar g\langle c^{\dag}\rangle\langle c\rangle]\langle qp+pq\rangle-2\gamma_{m}\langle p^{2}\rangle\nonumber\\& &+2\gamma_{m}(1+2n)\frac{m\hbar \omega_{m}}{2},\nonumber\\
\frac{d}{dt}\langle qp+pq\rangle&=&\frac{2}{m}\langle p^{2}\rangle+2(-m\omega_{m}^{2}-2\hbar g\langle c^{\dag}\rangle\langle c\rangle)\langle q^{2}\rangle\nonumber\\& &-\gamma_{m}\langle qp+pq\rangle,
\end{eqnarray}
in which the constant $2\gamma_{m}(1+2n)\frac{m\hbar \omega_{m}}{2}$ is due to the coupling of the membrane to the thermal environment, and $n=[e^{\frac{\hbar \omega_{m}}{k_{B}T}}-1]^{-1}$ is the mean phonon occupation
number of energy $\hbar \omega_{m}$ at temperature $T$, and where $k_{B}$ is the Boltzmann's constant. Note that the constant $(1+2n)\frac{m\hbar \omega_{m}}{2}$ is the mean value of the square of the momentum of the membrane.

We would solve Eq. (\ref{4}) under the assumption
that the coupling field is much stronger than the probe field. The steady state solution of Eq. (\ref{4}) then can be written as
\begin{equation}\label{5}
 \begin{array}{lcl}
 \left(
  \begin{array}{cccc}
   \langle c\rangle \\ \langle c^{\dag}\rangle\\  \langle q^{2}\rangle \\  \langle p^{2}\rangle\\ \langle qp+pq\rangle\\
  \end{array}
\right)=\left(
  \begin{array}{cccc}
   c_{0}\\ c_{0}^{*}\\ X_{0}\\ Y_{0}\\ Z_{0}\\
  \end{array}
\right)+\varepsilon_{p}e^{-i(\omega_{p}-\omega_{c})t}\left(
  \begin{array}{cccc}
    c_{+}\\ c_{-}^{*}\\ X_{+}\\ Y_{+}\\ Z_{+}\\
  \end{array}
\right)\\\hspace{0.7in}+\varepsilon_{p}^{*}e^{i(\omega_{p}-\omega_{c})t}\left(
  \begin{array}{cccc}
   c_{-}\\ c_{+}^{*}\\ X_{-}\\ Y_{-}\\ Z_{-}\\
  \end{array}
\right).
\end{array}
\end{equation}
The solution contains three components, which in the original frame oscillate at $\omega_{c}$, $\omega_{p}$, $2\omega_{c}-\omega_{p}$, respectively. Substituting Eq. (\ref{5}) into Eq. (\ref{4}), dropping those terms that contains the product of more than one small quantity, then equating coefficients of terms with the same frequency, respectively, we obtain
\begin{eqnarray}\label{6}
X_{0}&=&\frac{Y_{0}}{m^{2}\omega_{m}^{2}(1+2\alpha)},\nonumber\\
Y_{0}&=&(1+2n)\frac{m\hbar \omega_{m}}{2},\nonumber\\
c_{0}&=&\frac{\varepsilon_{c}}{\kappa+i\Delta},\nonumber\\
c_{+}&=&\frac{1}{d(\delta)}\{[\kappa-i(\Delta+\delta)](\gamma_{m}-i\delta)(\delta^{2}-4\omega_{m}^{2}+2i\gamma_{m}\delta\nonumber\\& &-8\alpha \omega_{m}^{2})-4i\alpha\beta\omega_{m}^{3}(2\gamma_{m}-i\delta)\},\nonumber\\
c_{-}&=&\frac{1}{d^{*}(\delta)}[-4i\alpha\beta \omega_{m}^{3}\frac{c_{0}^{2}}{|c_{0}|^{2}}(2\gamma_{m}+i\delta)],
\end{eqnarray}
where
\begin{eqnarray}\label{7}
\alpha&=&\hbar g |c_{0}|^{2}/(m \omega_{m}^{2}),\nonumber\\
\beta&=&gX_{0}/\omega_{m},\nonumber\\
\Delta&=&\omega_{0}-\omega_{c}+\beta\omega_{m},\nonumber\\
\delta&=&\omega_{p}-\omega_{c},\nonumber\\
d(\delta)&=&[\kappa+i(\Delta-\delta)][\kappa-i(\Delta+\delta)](\gamma_{m}-i\delta)\nonumber\\& &\times(\delta^{2}-4\omega_{m}^{2}+2i\gamma_{m}\delta-8\alpha \omega_{m}^{2})\nonumber\\& &+8\Delta\alpha\beta\omega_{m}^{3}(2\gamma_{m}-i\delta).
\end{eqnarray}
From Eqs. (\ref{6}) and (\ref{7}), we find that the cavity field at the probe frequency $\omega_{p}$ is related to the component $X_{0}$ of the mean-square amplitude of the motion of the membrane, which depends on the pump power and the temperature of the environment. And the coupling strength between the cavity field at the frequency $\omega_{p}$ and the membrane is affected by the quadratic coupling constant $g$ and the photon number $|c_{0}|^{2}$ in the cavity. Note that the parameter $\beta$ is a measure of the frequency shift of the cavity due to quadratic coupling. The parameter $\alpha$ is the ratio of the radiation pressure energy to the potential energy of the membrane.

Further, the output field can be derived by using the input-output relation
\begin{equation}\label{8}
\varepsilon_{out}(t)+\varepsilon_{p}e^{-i\delta t}+\varepsilon_{c}=2\kappa\langle
\tilde{c}\rangle.
\end{equation}
If we write $\varepsilon_{out}(t)$ as
\begin{equation}\label{9}
\varepsilon_{out}(t)=\varepsilon_{out0}+\varepsilon_{out+}\varepsilon_{p}e^{-i\delta t}+\varepsilon_{out-}\varepsilon_{p}^{*}e^{i\delta t},
\end{equation}
where $\varepsilon_{out0}$ is the response at the frequency $\omega_{c}$ of the coupling field, $\varepsilon_{out+}$ is the response at the frequency $\omega_{p}$ of the probe field, and $\varepsilon_{out-}$ is the response at the new frequency $2\omega_{c}-\omega_{p}$.
Combining Eqs. (\ref{8}) and (\ref{9}), we obtain
\begin{eqnarray}\label{10}
\varepsilon_{out0}&=&2\kappa c_{0}-\varepsilon_{c},\nonumber\\
\varepsilon_{out+}&=&2\kappa c_{+}-1,\nonumber\\
\varepsilon_{out-}&=&2\kappa c_{-}.
\end{eqnarray}
We examine the total output field at the frequency $\omega_{p}$ defined as $\varepsilon_{T}=\varepsilon_{out+}+1=2\kappa c_{+}$, so $\varepsilon_{T}$ is also affected by the pump power and the temperature of the environment. In the absence of the quadratic optomechnaical coupling ($g=0$), $\varepsilon_{T}$ is given by
\begin{equation}\label{11}
\varepsilon_{T}=\frac{2\kappa}{\kappa+i(\Delta-\delta)}.
\end{equation}

\begin{figure}[!h]
\begin{center}
\scalebox{0.65}{\includegraphics{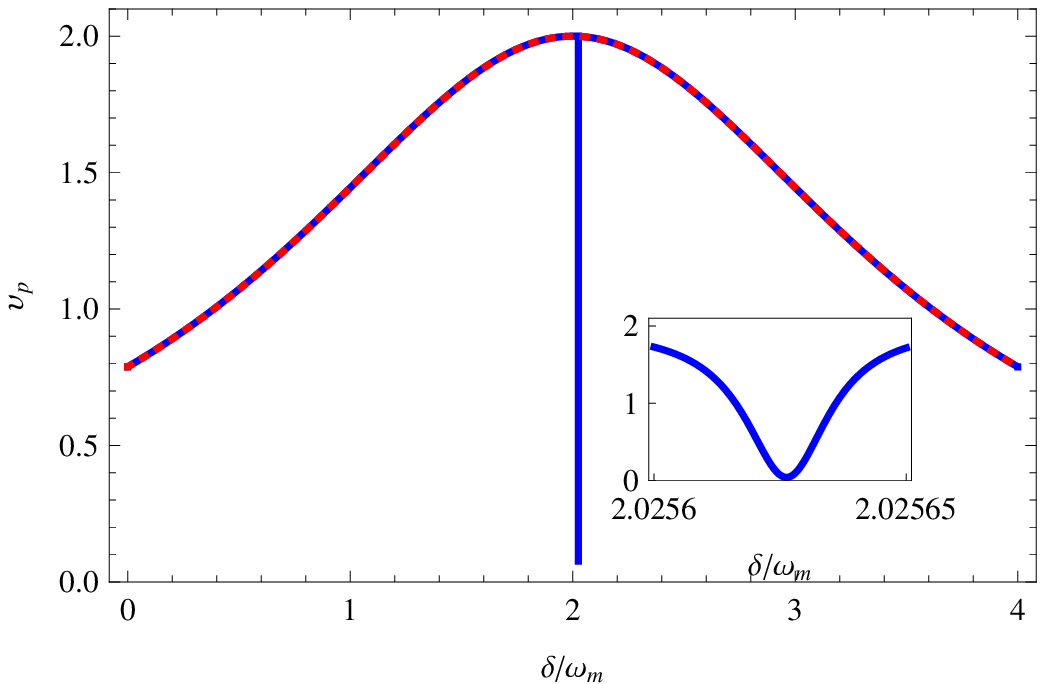}} \caption{\label{Fig2} (Color online) Quadrature of the output field $\upsilon_{p}$ as a function of the normalized frequency $\delta/\omega_{m}$ in the absence (red dotted line) and presence (blue solid line) of the quadratic coupling. Parameters: $\gamma_{m}=1$ s$^{-1}$, $r_{c}=0.42$, $\wp_{c}=20$ $\mu$W, and $T=20$ K. The inset zooms the EIT-like dip.}
\end{center}
\end{figure}

\begin{figure}[!h]
\begin{center}
\scalebox{0.65}{\includegraphics{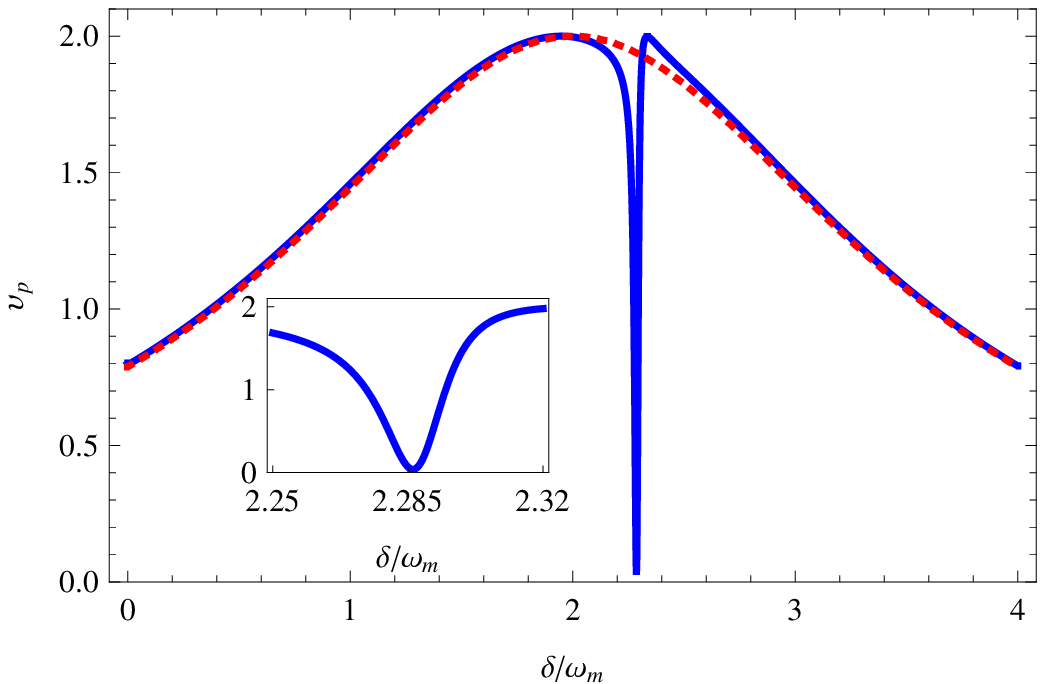}} \caption{\label{Fig3} (Color online) Quadrature of the output field $\upsilon_{p}$ as a function of the normalized frequency $\delta/\omega_{m}$ in the absence (red dotted line) and presence (blue solid line) of the quadratic coupling. Parameters: $\gamma_{m}=900$ s$^{-1}$, $r_{c}=0.999$, $\wp_{c}=10$ $\mu$W, and $T=100$ K. The inset zooms the EIT-like dip.}
\end{center}
\end{figure}

\begin{figure}[!h]
\begin{center}
\scalebox{0.65}{\includegraphics{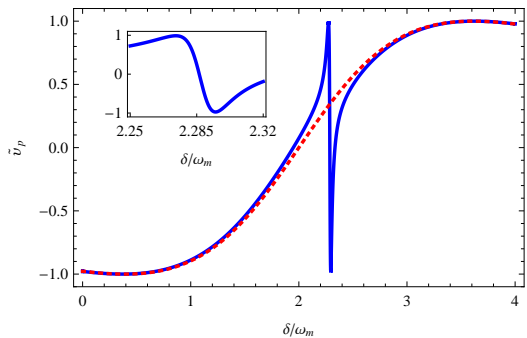}} \caption{\label{Fig4} (Color online) Quadrature of the output field $\tilde{\upsilon}_{p}$ as a function of the normalized frequency $\delta/\omega_{m}$ in the absence (red dotted line) and presence (blue solid line) of the quadratic coupling. Parameters: $\gamma_{m}=900$ s$^{-1}$, $r_{c}=0.999$, $\wp_{c}=10$ $\mu$W, and $T=100$ K. The inset zooms the EIT-like dip.}
\end{center}
\end{figure}

\section{EIT in the output field}
In this section, we calculate numerically the output field at the frequency $\omega_{p}$ to bring out the EIT-like phenomenon due to the interaction between the cavity field and the membrane which is quadratic dependence on the position of the membrane coupled quadratically to the cavity. For convenience, we write $\varepsilon_{T}$ as
\begin{equation}\label{12}
\varepsilon_{T}=\upsilon_{p}+i\tilde{\upsilon}_{p},
\end{equation}
The $\upsilon_{p}$ and $\tilde{\upsilon}_{p}$ give the inphase and out of phase quadratures of the output field. The quadratures can be measured via homodyne technique ~\cite{Walls}.

In order to explicitly demonstrate the possibility of EIT in quadratically coupled optomechanical systems we use parameters from Ref. ~\cite{Thompson}. This reference discusses many different possible scenarios for quadratic couplings. A later paper ~\cite{Sankey} gives experimental demonstration of how to achieve much larger quadratic couplings. We list the parameters used in numerical results. The wavelength of the coupling field $\lambda=\frac{2\pi c}{\omega_{c}}=532$ nm, the total cavity length $L=6.7$ cm, the frequency of the membrane $\omega_m=2\pi\times100$ kHz, the cavity finesse $F=6940$, the cavity decay rate $\kappa=\frac{\pi c}{2FL}=2\pi\times1.61\times10^{5}$ Hz, the decay rate of the membrane $\gamma_{m}=1$ s$^{-1}$, the mechanical quality factor $Q=\frac{\omega_{m}}{\gamma_{m}}=6.28\times10^5$, the membrane's reflectivity $r_{c}=0.42$, the coupling constant $g=8\pi^2c/(L\lambda^{2}\sqrt{2(1-r_{c})})=2\pi\times1.85\times10^{23}$ Hz/m$^{2}$, in which $c$ is the speed of light in vacuum. The pump power $\wp_{c}=20$ $\mu$W, and the temperature of the environment $T=20$ K. The mass of the membrane we use is $m=10^{-9}$g, which is less than that in Ref. ~\cite{Thompson}. In addition, we consider the two phonon resonance case $\Delta=2\omega_{m}$. It is good to compare the magnitude of the optomechanical coupling to the potential energy of the membrane. For coupling laser power of 20 $\mu$W; the parameter $\hbar g |c_{0}|^{2}$ at $\Delta=2\omega_{m}$ is 0.005 J/m$^{2}$; where as the parameter $m \omega_{m}^{2}$ is 0.4 J/m$^{2}$. Note that the ratio of $2\omega_{m}/\kappa$ is about 1.24 and thus these parameters are not quite in two phonon sideband resolved limit as the cavity finesse is not high enough.

Fig. ({\ref{2}) shows the phase quadrature $\upsilon_{p}$ as a function of the normalized frequency $\delta/\omega_{m}$ in the absence (red dotted line) and presence (blue solid line) of the optomechanical coupling, respectively. In the absence of the optomechanical coupling, from the red dotted line in Fig. ({\ref{2}), it is seen that $\upsilon_{p}$ have the standard absorption shape. However, in the presence of the optomechanical coupling, from the blue solid line in Fig. ({\ref{2}), one can clearly see an EIT-like dip in the phase quadrature $\upsilon_{p}$ when two phonon processes happen ($\delta\approx2\omega_{m}$). The position of the EIT-like dip is not exactly at $\delta=2\omega_{m}$ due to the term $8 \alpha \omega_{m}^{2}$ in $c_{+}$ and $d(\delta)$, in which $\alpha=0.013$. Note that the linewidth of the dip is extremely narrow due to $\gamma_{m}\ll\kappa$. The linewidth is about 10 Hz.

However, for other set of parameters, the EIT window becomes wider and more accessible to experiments. For the decay rate of the membrane $\gamma_{m}=900$ s$^{-1}$, the mechanical quality factor $Q=\frac{\omega_{m}}{\gamma_{m}}=698$, the membrane's reflectivity $r_{c}=0.999$, the coupling constant $g=2\pi\times4.44\times10^{24}$ Hz/m$^{2}$, the pump power $\wp_{c}=10$ $\mu$W, and the temperature of the environment $T=100$ K, the phase quadratures $\upsilon_{p}$ and $\tilde{\upsilon}_{p}$ as a function of the normalized frequency $\delta/\omega_{m}$ in the absence (red dotted line) and presence (blue solid line) of the optomechanical coupling are given in Figs. (\ref{3}) and (\ref{4}). From the blue solid line in Fig. (\ref{3}), we can see the linewidth of the EIT-like dip is about $0.02\omega_{m}=$12566 Hz, thus the EIT-like dip with quadratic membranes should be detectable rather easily. We also find the position of the EIT-like dip is at $\delta\approx2.285\omega_{m}$, away from $\delta=2\omega_{m}$, which is due to the large value of the parameter $8 \alpha \omega_{m}^{2}$, where $\alpha=0.155$. Moreover, in the the case without the optomechanical coupling, from the red dotted line in Fig. ({\ref{4}), it is seen that $\tilde{\upsilon}_{p}$ has standard dispersion shape. But in the case with the optomechanical coupling, from the blue solid line in Fig. ({\ref{4}), we can see the phase quadrature $\tilde{\upsilon}_{p}$ exhibits abnormal dispersion.

\section{conclusions}
In conclusion we have shown how EIT-like effects can arise in two phonon processes in optomechanical systems. Our finding is a new paradigm for EIT as what plays the role of atomic coherence is zero for quadratically coupled systems. The basic quantity leading to EIT in our system is the fluctuation in the displacement of the membrane. Interestingly enough the EIT-like behavior can occur at very low coupling powers like tens of microwatts.

We gratefully acknowledge support from the NSF Grant No. PHYS 0653494.

\end{document}